\documentclass{article}
\usepackage{moriond}
\usepackage{graphicx}
\usepackage{natbib}
\begin{document}
\vspace*{4cm}
\title{XMM-NEWTON OBSERVATIONS OF GALAXY CLUSTERS: THE RADIAL TEMPERATURE PROFILE OF A2163}

\author{ G.W. PRATT$^1$, M. ARNAUD$^1$ and N. AGHANIM$^2$}

\address{$^1$ CEA/Saclay, Service d'Astrophysique, L'Orme des Merisiers, B\^{a}t. 709, 91191 Gif-sur-Yvette Cedex, France \\
         $^2$ IAS-CNRS, Universit\'{e} Paris Sud, B\^{a}t. 121, 91405 Orsay Cedex, France}

\maketitle\abstracts{XMM-Newton, with its high throughput and excellent spatial and spectral resolution, is an ideal instrument for spectro-imaging observations of clusters. Presented here is an XMM-Newton mosaic observation of A2163, from which a new radial temperature profile is derived. The issue of background subtraction is addressed in some detail as a proper treatment of the background is essential for the derivation of correct temperatures, and thus, of the correct total cluster mass. }


\section{Introduction}

The main baryonic component of galaxy clusters is the hot X-ray emitting gas of the intracluster medium. Provided the gas density distribution (obtained from the surface brightness profile), overall temperature, and temperature structure of the gas are known, the gravitational potential can be measured in the assumption of hydrostatic equilibrium. Radial temperature profiles of the intracluster gas are thus an essential tool for the determination of the total gravitational mass contained within clusters. Most early approaches assumed an isothermal gas temperature distribution for the derivation of the mass profile (e.g., for A2163, the {\it ROSAT\// GINGA\/} analysis of Elbaz, Arnaud \& B\"{o}hringer 1995). However, since the advent of {\it ASCA} and more recently {\it BeppoSAX\/}, allowing spatially-resolved spectroscopy, this assumption has been called into question.

Numerical simulations suggest that the temperature profiles of clusters are flat out to $\sim 0.5$ of the virial radius ($r_v$), after which there is a gentle decline to about 20\%-50\% of the central temperature at $r_v$ itself (Evrard, Metzler \& Navarro 1996; Navarro, Frenk \& White 1995b; Frenk et al. 1999). From the observational standpoint, though, the existence of a decline at large radii ($r > 0.5r_v$) is still an open question. This is because, even with large samples, the number of clusters where the temperature is derived beyond $0.5r_v$ is in general very small. (For example, in the sample of 106 clusters analysed by White [2000], only 4 have temperatures derived beyond $0.5r_v$.)

Three analyses of large samples of clusters serve to illustrate the current status of the debate. Markevitch et al. (1998) studied {\it ASCA\/} observations of 30 clusters and found that in general temperature profiles declined significantly with radius, and proposed a universal profile based on this evidence. However, White (2000) also studied 106 {\it ASCA\/} observations, applying a different PSF correction technique from that of Markevitch et al. and adding a cooling flow treatment to the modelling; of this sample, 90\% had profiles that were consistent with isothermality. Finally, Irwin \& Bregman (2000) studied 11 clusters observed with {\it BeppoSAX\/}, and found that the temperature profiles were generally flat. None of these samples is independent, which suggests that different data treatments and modelling may have an effect on the final form of the temperature profiles.

A2163 is an exceptional cluster in many ways, hot, massive and luminous\footnote{$T = 14.6^{+0.9}_{-0.8}$ keV, $M_{\rm grav}(r < 4.6$ Mpc$)$ = $4.6^{+0.4}_{-1.5} \times 10^{15}$ $M_\odot$, $L_X[2-10]$ keV = $6.0 \times 10^{45}$ ergs s$^{-1}$, redshift $z = 0.201$; all values from Elbaz et al. 1995}, but crucially it is one of a select few where it is possible to measure the temperature beyond $0.5 r_v$, and so provides an observational test of the predictions from numerical simulations. Several radial temperature profiles of this object exist. {\it ASCA\/} observations by Markevitch et al. (1996) showed a very significant drop in temperature at large radii ($\sim 50$\%), a result which has been recently supported with preliminary results from {\it Chandra} observations (Markevitch et al. 2000). However, White (2000) found a flat temperature profile from the {\it ASCA\/}, while Irwin \& Bregman (2000) find that the profile may {\it rise} slightly from analysis of {\it BeppoSAX\/} data.

A2163 was observed by XMM-Newton as part of a GT programme involving a collaboration between the CEA Service d'Astrophysique, IAS Orsay, and the Universities of Leicester and Birmingham; presented here is a radial temperature profile derived from EPIC/MOS data. 


\section{Data analysis}


\subsection{Observations, data processing and vignetting treatment}

The observation of A2163 consists of five pointings of XMM-Newton, all obtained in the EPIC Extended Full Frame Mode between revolutions 0132 and 0137. 
The SASv5.0.1 processing task {\it emchain} was used to generate calibrated events files, using the option to detect any remaining bad pixels. A check after processing revealed that these pixels were indeed successfully removed.

Vignetting is the reduction in the effective area of the telescope at a given energy, and is dependent on the detected position of the photon in the field of view. This effect must accounted for in any analysis, but especially so in the case of extended sources such as clusters of galaxies.
The photon weighting method of Arnaud et al. (2001) was used to correct for the vignetting effect.










\subsection{Treatment of the background}

The XMM-Newton background is energy dependent and consists of several components. Source spectra must be fully corrected for the background, as inadequate consideration can lead to systematic errors in any parameters resulting from the modelling. The total background can be considered in terms of contributions from:

\begin{enumerate}

\item a component induced by soft protons from solar flares;

\item a hard X-ray component ($E > 1.5$ keV), consisting of 

	\begin{itemize}

	\item a cosmic hard X-ray background, which is vignetted,

	\item and a cosmic ray-induced (particle) background, which mimics the hard X-ray background, is the dominant hard component, and is not vignetted;

	\end{itemize}

\item a cosmic soft X-ray component ($E < 1.5$ keV), which is variable from pointing to pointing (e.g., Snowden et al. 1997), and is vignetted;

\item and an instrumental component, e.g., fluorescence lines (Al, Si) from the material of the telescope itself.

\end{enumerate} 

The background induced by solar soft proton flares (1) is by now well-known, affecting both {\it XMM\/} and {\it Chandra} observations. The temporal and flux-variability of the flares lead to similar variability in the spectrum, and so these periods cannot be accounted for in the usual way. These data must be removed. Accordingly, all frames with a count rate greater than 15 ct/100s in the [10-12] keV band were excised from the calibrated event files of all pointings. The remaining background is dominated by the cosmic soft X-ray component at low energy and the cosmic ray-induced component at energies above $\sim 1.5$ keV.


In order to estimate this remaining background, the event files combining several high galactic latitude pointings generated by D. Lumb were used, filtered as above for periods of high soft proton flux, with bright sources removed. Hereafter this will be referred to as the ``blank-sky'' background. Note that it is essential to apply the same soft proton filtering criteria to both the source and blank-sky background.

After filtering the calibrated event files for the contribution of the soft proton flux, a merged event list corresponding to the mosaic of MOS1 and MOS2 pointings was then generated. At present the quality of the vignetting calibration data at large off-axis angles degrades the statistical quality of detections in these regions. To maximise the signal to noise only events within the central $12^{\prime}$ of each pointing were thus considered; this can be regarded as a conservative criterion. 

It is known that the cosmic ray-induced hard background (2a) changes slightly in the field of view, so it is important to consider the same extraction regions in the detector for both the source and background. A mosaic of blank-sky pointings, fully equivalent to the mosaic of A2163, was thus generated by transforming the sky coordinates of the blank-sky events file into the sky coordinates of each of the five pointings of the cluster.

\begin{figure}[h]
\begin{centering}
\includegraphics[scale=0.6,angle=0,keepaspectratio]{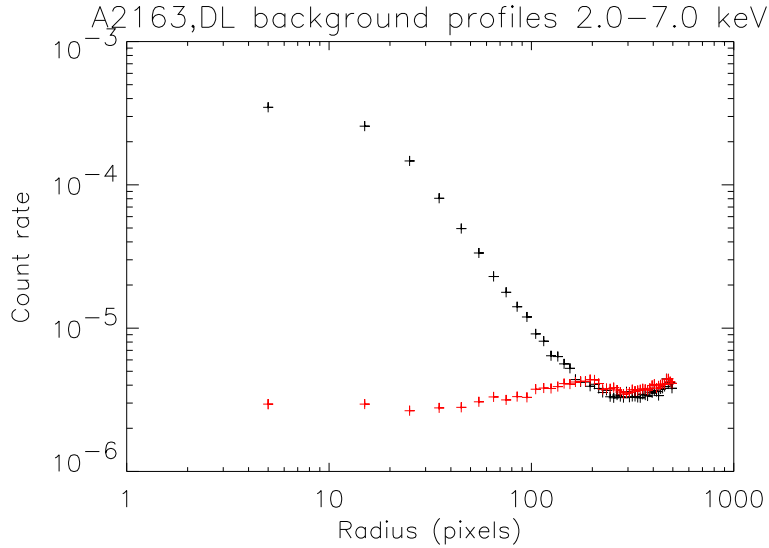}
\caption{{\footnotesize A comparison of the surface brightness profiles, in the band [2-7] keV, of the mosaic of A2163 (black) and the blank-sky background mosaic (red). Here 1 pixel = $3.3^{\prime\prime}$}}\label{fig:sbprof1}
\end{centering}
\end{figure}

\begin{figure}[h]
\begin{centering}
\includegraphics[scale=0.4,angle=270,keepaspectratio]{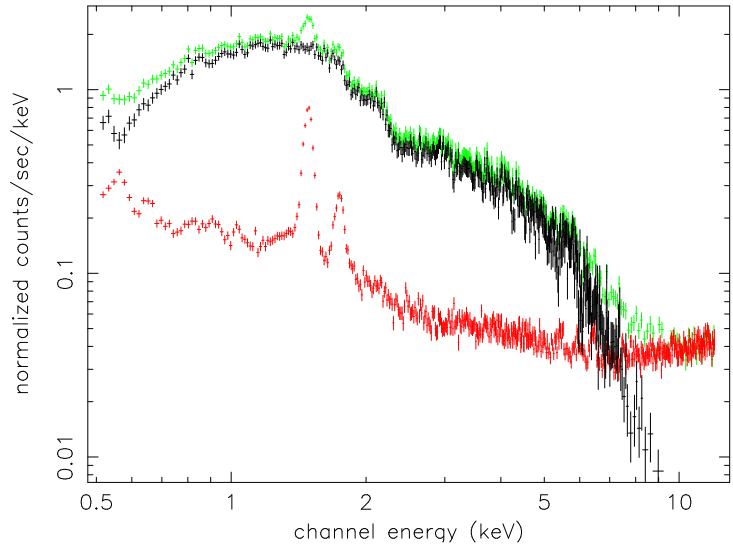}
\caption{{\footnotesize Subtraction of the hard X-ray background contribution from the global $(0^{\prime}.0 - 8^{\prime}.8)$ spectrum of A2163. The raw spectrum is plotted in green, the (normalised) background spectrum derived from the blank-sky event files is in red, and the resulting background-subracted spectrum is in black}}\label{fig:bgsub1}
\end{centering}
\end{figure}

To normalise the blank-sky background, a surface brightness profile was produced in the band [2-7] keV for both the cluster and blank-sky mosaics, centred on the peak of the X-ray emission from A2163. These are shown in Figure~\ref{fig:sbprof1}: the difference in flux in the outer regions yields a precise estimate of the normalisation. This treatment allows a proper evaluation of the hard X-ray background contribution (2a,b), and as the instrumental background (4) appears in both the cluster and blank-sky observations, this component is also removed. In Figure~\ref{fig:bgsub1} the subtraction of the normalised blank-sky background is illustrated for the global spectrum of A2163.

\begin{figure}[h]
\begin{centering}
\includegraphics[scale=0.6,angle=0,keepaspectratio]{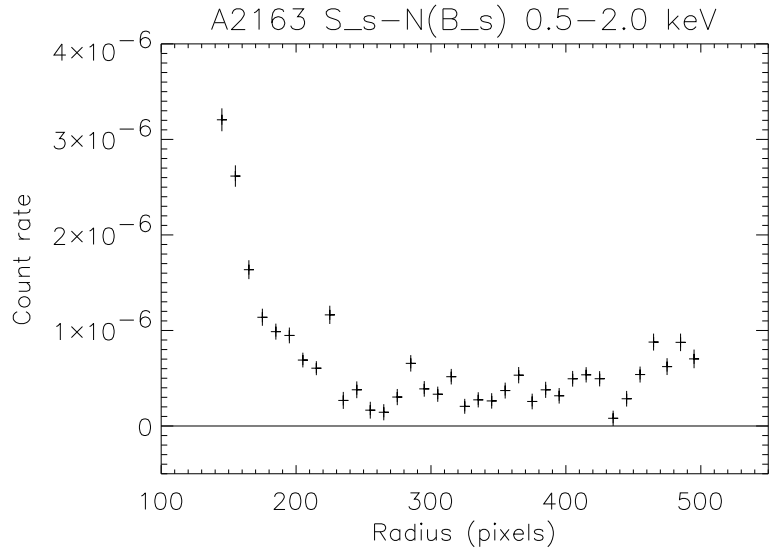}
\caption{{\footnotesize The surface brightness profile of the outer regions of the source mosaic in the band [0.5-2] keV, after subtraction of the blank-sky background normalised in the [2-7] keV band. See text for full details}}\label{fig:sbprof2}
\end{centering}
\end{figure}

Lastly, it is necessary to consider the residual background due to soft X-ray diffuse emission (3), which varies with position on the sky. For A2163, this is especially important as the cluster lies in a region of the sky affected by significant amounts of such emission (see Snowden et al. 1997). Figure~\ref{fig:sbprof2} shows the surface brightness profile of the outer regions of the A2163 mosaic in the band [0.5-2] keV, after subtraction of the blank-sky background normalised as above.

\begin{figure}[h]
\begin{centering}
\includegraphics[scale=0.4,angle=270,keepaspectratio]{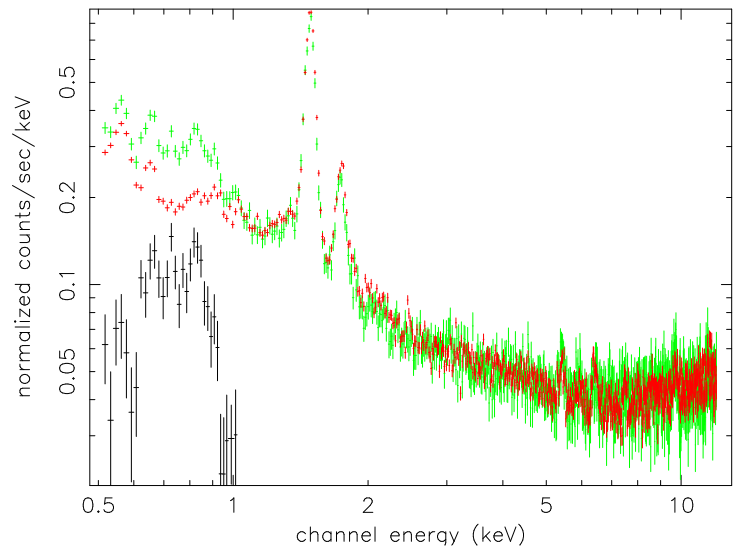}
\caption{{\footnotesize The spectrum of the external ($12^{\prime} - 15^\prime$) region of the blank-sky mosaic (red) is subtracted from the spectrum of the same region from the A2163 mosaic (green), to give the resulting residual soft X-ray background spectrum (black)}}\label{fig:bgsub2}
\end{centering}
\end{figure}

\begin{figure}[h]
\begin{centering}
\includegraphics[scale=0.4,angle=270,keepaspectratio]{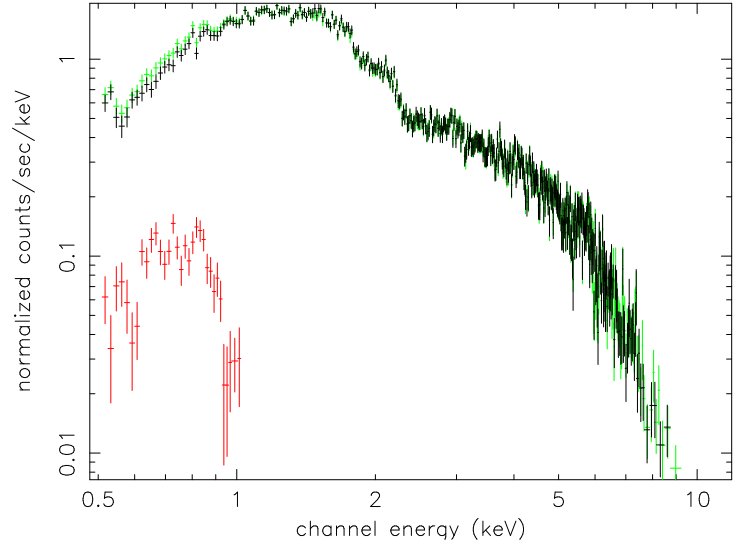}
\caption{{\footnotesize Subtraction of the soft X-ray background component (red) from the blank-sky background-subtracted spectrum (green) gives the final global spectrum of A2163 (black)}}\label{fig:bgsub3}
\end{centering}
\end{figure}

 A clear excess of flux, associated with the soft X-ray background, is seen. To remove this last component, a spectrum was extracted from an annular region of the mosaic situated between $12^{\prime}$ and $25^{\prime}$ from the peak of the emission of A2163 a region which is external to the emission from the cluster. A spectrum was likewise extracted from the same region of the blank-sky mosaic. The blank-sky spectrum was then subtracted from the cluster spectrum, thus giving the residual spectrum of the soft X-ray background. This was then directly subtracted from all subsequent source spectra after normalisation to the size of the extraction region. Figures~\ref{fig:bgsub2} and~\ref{fig:bgsub3} illustrate the generation of the soft X-ray background spectrum and its subsequent subtraction from the global spectrum of A2163.

\section{The temperature and mass profiles}

\begin{figure}[h]
\begin{centering}
\includegraphics[scale=0.46,angle=0,keepaspectratio]{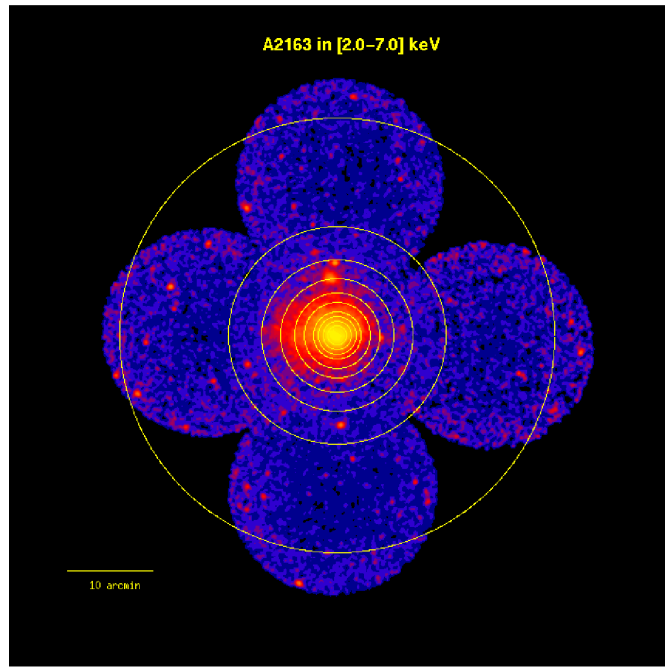}
\caption{{\footnotesize The A2163 mosaic showing the annular extraction regions for the radial temperature profile}}\label{fig:mosaic}
\end{centering}
\end{figure}

Figure~\ref{fig:mosaic} shows the mosaic image of A2163. Overlaid are the annular extraction regions, centred on the peak of the X-ray emission of the cluster, from which the radial temperature profile was produced after removal of bright point sources. The final annulus extends between $6'.6$ and $8'.8$ from the cluster core. The outer annulus is the extraction region chosen for the determination of the residual soft-X-ray background (see above). 

The annular spectra were then fitted with an absorbed thermal plasma model in the $0.5-10.0$ keV band. The usual procedure is to take the value of the galactic absorption as deduced from 21cm measurements, and fit the spectra with this parameter fixed. However, the value of the absorption to A2163 is higher than the galactic value, and was found to be $N_H = 1.65 \times 10^{21}$ cm$^{-2}$ in the analysis of the global {\it ROSAT\/} PSPC spectrum by Elbaz et al. (1995). The spectra were thus fitted with the absorption column fixed at this value (see also discussion below).

\begin{figure}[]
\begin{centering}
\includegraphics[scale=0.4,angle=270,keepaspectratio]{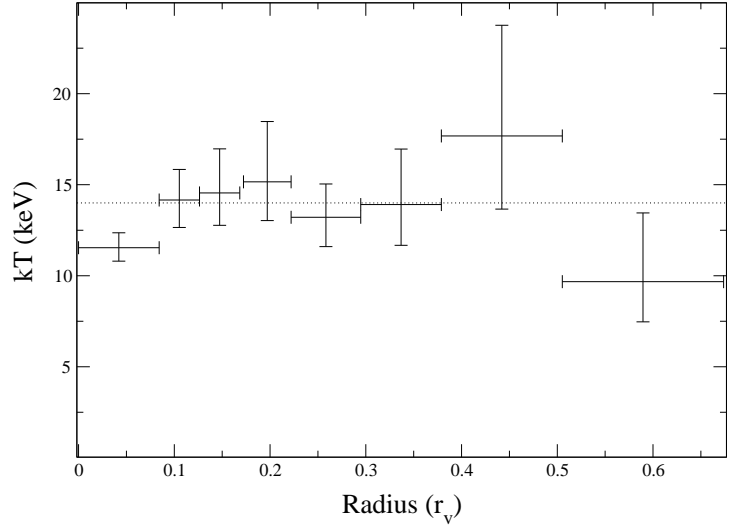}
\caption{{\footnotesize The XMM-Newton projected radial temperature profile of A2163 in units of the virial radius. The dotted line is the temperature of the global [0'.0 - 8'.8] spectrum (14.0 keV). Errors are 90\% confidence level}}\label{fig:rtprof}
\end{centering}
\end{figure}

The final projected radial temperature profile, in units of the virial radius, is shown in Figure~\ref{fig:rtprof}.

\vspace{0.5cm}

\begin{figure}[h]
\begin{centering}
\includegraphics[scale=0.5,angle=0,keepaspectratio]{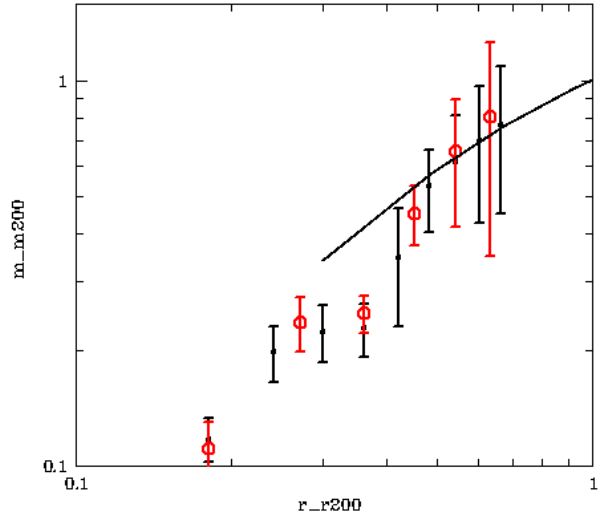}
\caption{{\footnotesize A preliminary total mass profile, for sampling windows of 200 (black) and 300 (red) kpc. The solid line is the result from the numerical simulations of Evrard et al. (1996) }}\label{fig:massprof}
\end{centering}
\end{figure}

A preliminary total mass profile was then produced using the Monte-Carlo method of Neumann \& B\"{o}hringer (1995), for which the new temperature data were used as input. The result, for two sampling windows\footnote{See Neumann \& B\"{o}hringer (1995) for an explanation} of 200 and 300 kpc, is shown  compared to the scaled profile from the simulations of Evrard et al. (1996) in Figure~\ref{fig:massprof}. The total mass found, $M(r \leq 2.2$ Mpc) $ = 2.8-3.0 \times 10^{15}$ M$_\odot$.

\newpage
\section{Discussion}

\subsection{The effect of background subtraction on the temperature profile}

{\it If the soft X-ray background is neglected}\\

\begin{figure}[h]
\begin{centering}
\includegraphics[scale=0.4,angle=270,keepaspectratio]{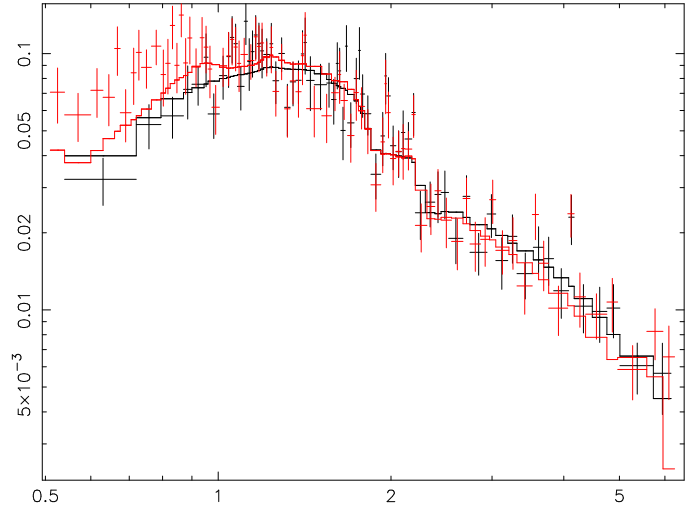}
\caption{{\footnotesize Spectra of the last annulus fitted with an absorbed thermal model with $N_H = 1.65 \times 10^{21}$ cm$^{-2}$ fixed. The temperature of the spectrum uncorrected for the soft X-ray background component (red) is cooler, at $kT = 5.89$ keV, than the corrected spectrum ($kT = 9.68$ keV, and the fit is considerably worse $\chi^2 = 1.43$ vs. 1.07. The fit is noticeably worse at soft energies }}\label{fig:lastancomp}
\end{centering}
\end{figure}

The background becomes more noticeable in the outer regions, where the cluster emission rapidly drops off. If the soft X-ray background component is not removed, the greatest effect will be in the last ($6'.6 - 8'.8$) annulus. The spectrum of this annulus, {\it with} and {\it without} subtraction of the soft component, was fitted with an absorbed thermal model with the column fixed at the {\it ROSAT\/} value of $N_H = 1.65 \times 10^{21}$ cm$^{-2}$. The temperature of the spectrum uncorrected for the soft component is lower, and the fit is worse, particularly at soft energies. Moreover, the difference in the derived temperatures is considerable- $kT = 5.8$ keV (uncorrected), $kT = 9.7$ keV (corrected). The fits are illustrated in Figure~\ref{fig:lastancomp}.

The temperature profile as a whole displays a much steeper drop in the temperature at large radii, if the soft component is not subtracted and the absorption is fixed at the {\it ROSAT\/} value.



\vspace{1cm}
\noindent{\it If the absorption is a free parameter}\\

\begin{figure}[h]
\begin{centering}
\includegraphics[scale=0.4,angle=270,keepaspectratio]{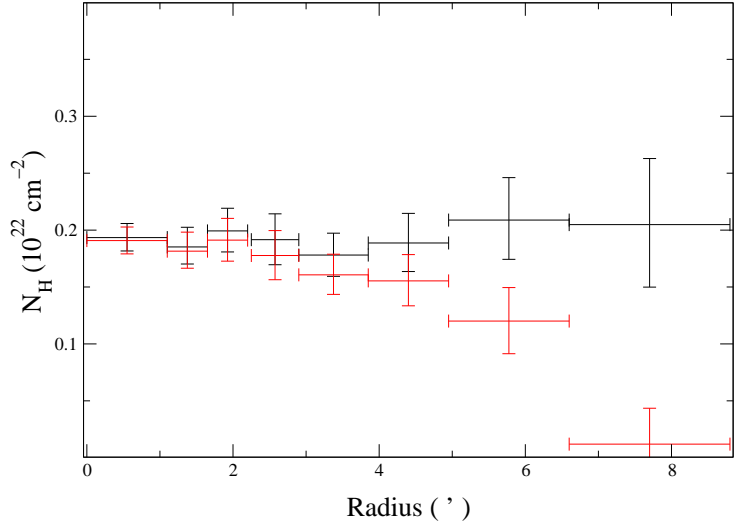}
\caption{{\footnotesize The value of $N_H$ for each annulus for spectra fully corrected for the background subtraction (black), and for spectra with all but the residual soft X-ray background component removed (red). See text for details }}\label{fig:abscomp}
\end{centering}
\end{figure}

The value of the absorption column derived for each annulus is also an independent check of the background subtraction procedure. The {\it ROSAT\/} PSPC map of the region in the vicinity of A2163 shows no variation within 5\% (Markevitch \& Vikhlinin 2001), so if background subtraction is correct, the value of the absorption column should be the same for each annulus. Modelling software will interpret any residual soft background contribution as a decrease in the $N_H$ with radius, as this component increases flux at low energies and is interpreted by the software as a decrease in $N_H$. Fitting annular spectra both corrected and uncorrected for the soft X-ray component with the absorption as a free parameter, it is found that the value of the absorption is constant for each fully corrected annulus, but decreases with radius for uncorrected spectra, as expected. At the cluster centre, where the background has least effect, the value of the absorption for both the corrected and uncorrected spectra is consistent with the {\it ROSAT\/} value; in the last annulus, the absorption of the corrected spectrum is still consistent, while for the uncorrected spectrum the derived absorption is $\sim 7$ times less. A comparison of the $N_H$ values is shown in Figure~\ref{fig:abscomp}.

\vspace{0.5cm}

In modelling of X-ray spectra, there is a complex interplay between the absorption and the temperature. The temperature derived for a spectrum can increase with decreasing $N_H$ as a hard spectrum can mimic a low absorption simply because there is less flux at low energies. Thus, if the absorption is left as a free parameter and the soft X-ray background component is not removed, the temperature profile {\it rises} towards the outer regions of the cluster because the derived $N_H$ decreases.

Finally, note that the value for the absorption derived from the fits to the fully corrected spectra is slightly higher than that derived from the {\it ROSAT\/} analysis, an effect which is probably the result of residual uncertainties in the {\it XMM-Newton} response files. A higher (fixed) absorption lowers the derived temperatures by a constant factor for each radial bin, and so does not change the form of the profile.


\subsection{The radial temperature and mass profiles}

When the background is correctly subtracted, the temperature profile of A2163 is consistent with being flat out to $0.5r_v$, after which there is evidence for a slight decline of $\sim 30_{-16}^{+26}$\% of the global temperature. The decline is not as sharp as that found by Markevitch et al. ($\sim 50$\%) but the temperature found here for the outer region is marginally consistent with that of Markevitch et al. within the errors. There is no clear evidence for a decline in the {\it BeppoSAX\/} study by Irwin \& Bregman (2000), and it may even rise, but their errors are large.

With regard to the preliminary mass profile, it is clear that in the outer regions the data agree very well with the Evrard et al. (1996) predictions. Temperature maps of A2163 show that the inner regions are probably not relaxed (Markevitch et al. 2000, Bourdin et al. this conf.), which is likely the reason for the divergence of the profiles below $\sim 0.4r_v$.


\section{Conclusions}

It has been found that the proper treatment and subtraction of all background components is {\it essential}, especially in the outer regions of clusters where the background becomes steadily more dominant. The residual soft X-ray component must be corrected for.

\begin{itemize}

\item If the soft X-ray component is neglected and the absorption is a free parameter to model fits, the derived $N_H$ decreases and the temperature profile rises towards the outer regions.

\item If the soft X-ray component is neglected and the absorption is fixed, the profile declines sharply in the outer regions.

\item If the soft X-ray component is corrected for, the profile declines gently in the outer regions. The results are fully consistent whether the absorption is fixed or free.

\end{itemize}

 The flat radial temperature profile out to $r_v/2$, followed by a slight decline out to $\sim 0.7r_v$, found from the mosaic observations presented here, is in agreement with the predictions from numerical simulations. A preliminary mass profile also agrees well with numerical simulations, especially in the outer regions of the cluster. The disagreement below $\sim 0.4r_v$ is likely due to this region being in a non relaxed state.

Future work involves the addition of the EPIC/pn data and the use of the mosaic to probe the intrinsic temperature structure of the cluster. A more detailed discussion will be presented in a forthcoming paper.


\end{document}